\documentclass[letterpaper,dvipsnames]{article}
\pdfoutput=1
\usepackage{jheppub}
\usepackage{amsmath}
\usepackage{graphicx}

\usepackage{array}
\usepackage{xcolor}

\usepackage{slashed}
\usepackage{afterpage}
\usepackage{appendix}
\usepackage{multirow}
\usepackage{float}
\usepackage{mathtools}
\usepackage{verbatim}
\mathtoolsset{centercolon}

\usepackage{multirow}

\newcommand{\CP}{{\sf CP}}
\newcommand{\eV}{\ {\rm eV}}

\title{Searching for axion forces with precision precession in storage rings}
\author[1]{Prateek Agrawal,}
\author[2]{David E. Kaplan,}
\author[3]{On Kim,}
\author[2]{Surjeet Rajendran}
\author[1]{and Mario Reig}

\affiliation[1]{Rudolf Peierls Centre for Theoretical Physics, University of Oxford, Parks Road, Oxford OX1 3PU, United Kingdom}
\affiliation[2]{Department of Physics and Astronomy, Johns Hopkins University, Baltimore, MD 21218, USA}
\affiliation[3]{Department of Physics and Astronomy, University of Mississippi, University, MS 38677, USA}

\abstract{We consider different types of storage rings as precision probes of axion-mediated monopole-dipole forces. We show that current and planned experiments aiming to measure magnetic and electric dipole moments of protons, muons and electrons very precisely may explore new parts of the parameter space beyond existing laboratory bounds and, in some cases, beyond astrophysical constraints. Remarkably, a light axion coupled to muons may explain the FNAL/BNL $(g-2)_\mu$ anomaly as an environmental effect -- the coherent axion field generated by the earth nucleons induces an extra contribution to the anomalous precession frequency of the muon explaining the discrepancy with respect to the SM prediction.}

\begin{document}

\maketitle

\section{Introduction}
The Standard Model (SM) has been remarkably successful in describing interactions of matter at the shortest distance scales probed so far. There remains, however, the intriguing possibility of additional light degrees of freedom that are coupled feebly with the SM, and can produce tiny but macroscopic effects such as fifth forces. High precision experiments are ideally poised to detect such small effects, potentially unlocking the key to mysteries of the dark sectors.

Axions are compelling candidates for such new physics, motivated from the low energy point of view by the strong {\CP} problem \cite{Weinberg:1977ma,Wilczek:1977pj,Peccei:1977hh} and the dark matter puzzle \cite{Preskill:1982cy,Abbott:1982af,Dine:1982ah}. They are naturally light due to their nature as pseudo-Nambu Goldstone bosons, and are ubiquitously present in string compactifications \cite{Svrcek:2006yi,Arvanitaki:2009fg}. From the point of view of this work, axions (or axion-like particles) exemplify very light spinless fields that couple to the Standard Model matter and mediate long-range forces. 

The shift symmetry that protects the axion mass also prevents it from mediating a Coulombic $1/r^2$ force. A spin-dependent force is consistent with the shift symmetry, which is a dipole force that is much more challenging observationally. 
The shift symmetry of the axion is broken by its mass, and its couplings can also inherit a small shift-symmetry breaking effect. In this case, there are extremely weak monopole-monopole forces, as well as monopole-dipole and dipole-dipole forces~\cite{Moody:1984ba}.
We will focus on the monopole-dipole forces for axions.  In an axion gradient background, the dipole interaction with fermions appears as an effective background magnetic field, $B_{\rm axion} \sim m_f \nabla a / f_a$, which generates precession of spins\footnote{See \cite{Barbieri:1985cp,Vorobev:1989hb} for early,  related ideas.}.

Spin precession due to the interaction with an axion gradient has been used previously to look for axion dark matter (DM)~\cite{Graham:2013gfa} and for sourced axions~\cite{Arvanitaki:2014dfa}. Recently, storage rings have been proposed to look for light axion DM and dark energy (DE)~\cite{Graham:2020kai} (see also \cite{Smorra:2019qfx,Kim:2021eye,JEDI:2022hxa,Stephenson:2020jzx}).
In this paper we will consider sensitivity of storage rings to axion gradients sourced by the earth (which is sizeable for light axions $m_a \lesssim 10^{-10} \eV$) or test masses through the monopole interaction with nucleons.

The muon $g-2$ experiment is one important example. Indeed, we will show that the effect on the muon spin precession can almost be as big as the discrepancy between theory and experiment, with a mild tension with astrophysical limits~\cite{Bollig:2020xdr,Caputo:2021rux}. Independent of the experimental anomaly, our signal highlights the motivation for going to higher precision experiments beyond the precision of the theoretical calculations of muon $g-2$.

Proton storage rings can potentially achieve exquisite precision on the measurement of spin precession~\cite{Alexander:2022rmq,CPEDM:2019nwp,PhysRevD.105.032001}, making them ideal targets for measuring these couplings. The recent advances in measurement of electron EDMs (see \cite{Alarcon:2022ero} for a recent review) may also be leveraged to provide the strongest lab-based limits on axion-electron couplings.

We review the basic setup and the signal in section~\ref{sec:setup}. We then study different types of storage ring experiments and estimate their sensitivity to long-range axion forces in sections~\ref{sec:muon},~\ref{sec:protons},~\ref{sec:electrons}. We conclude in section~\ref{sec:conclusion}.

\section{Axion forces and spin precession experiments}
\label{sec:setup}
We consider models of the type
\begin{equation}\label{eq:lag}
    \mathcal{L}\supset g_s \phi \Bar{N}N+c_\psi\frac{\partial_\mu \phi}{f_\phi}\Bar{\psi}\gamma^\mu\gamma^5\psi\,,
\end{equation}
where $g_s$ is the CP violating scalar\footnote{The UV nature of the scalar coupling determines if it is Equivalence Principle (EP) violating, or not. For example when $g_s$ is not exactly proportional to the nucleon mass but to light quark masses times a CP violating phase, or is generated through the mixing with the Higgs, its EP violating behaviour is below $O(1\%)$ \cite{Graham:2015ifn}.} coupling to nucleons, $N$, while $g_p^\psi=\frac{c_\psi m_\psi}{f_\phi}$ determines the CP conserving dipole coupling to the fermion $\psi$. We work with the convention that $c_\psi=1$ absorbing the dependence in $f_\phi$. We remain agnostic to particular UV completions giving rise to these couplings\footnote{We acknowledge that for a QCD axion the scalar coupling, which is CP violating, is roughly given by $g_s\sim \theta_\text{eff}\frac{m_N}{f_a}$. In BSM extensions with additional sources of CP violation, the induced scalar coupling may be larger (see \cite{Dekens:2022gha} for a comprehensive study).} and instead focus on their phenomenology at spin precession experiments using storage rings. 

In the non-relativistic limit\footnote{See \cite{Silenko:2006er,Silenko:2021qgc} for a derivation using the Foldy-Wouthuysen transformation. } this leads to the well-known interaction Hamiltonian \cite{Barbieri:1985cp,Vorobev:1989hb}:
\begin{equation}\label{eq:monopole-dipole_Hamiltonian}
    H_{\phi}=-\frac{1}{f_a}\mathbf{\nabla} \phi \cdot \mathbf{S}\,,
\end{equation}which leads to the monopole-dipole potential between a particle sourcing the coherent axion field and a spin:
\begin{align}
    V(r)
    &=
    \frac{g_s g^\psi_p}{8\pi m_\psi}
    \left(
    \frac{1}{\lambda_\phi r}+\frac{1}{r^2}
    \right)
    e^{-m_\phi r}\mathbf{S}\cdot \hat{r}\,,
\end{align}
where $m_\phi$ is the mass of the axion, and $\lambda_\phi\sim m_\phi^{-1}$ is its associated wavelength.

The Yukawa-like potential above has an interesting distance dependence. Let us assume that a spin is located at a distance $d$ with respect to a given test mass. The test mass, of size $D$, has  $N_N\sim n_N D^3$ nucleons, where $n_N$ is the number density of nucleons. Obviously for $D,d\gg \lambda_\phi$, the nucleons do not produce a sizeable coherent axion field affecting the spin -- the exponential suppression makes the force very weak at distances $r>\lambda_\phi$. As $\lambda_\phi$ ($m_\phi$) increases (decreases), so does the reach of the classical axion field. For $D\gg\lambda_\phi>d$ the number of nucleons that contribute non-negligibly to the potential on the spin grows as $N_N\sim n_N\lambda_\phi^3$. Therefore, in this regime, the total monopole-dipole potential, $V_T(r)=N_NV(r)$, on a spin grows linearly with $\lambda_\phi$. Finally, for $\lambda_\phi\gg D,d$, that is for axion wavelengths much larger than any other scale in the problem, all the nucleons of the test mass effectively produce a potential on the spin and the signal is constant with the axion mass. This dependence will be important when estimating the effect of the axion gradient generated by the earth on a detector made of polarized spins.
 
The temporal evolution of the spin (at rest) is described by the equation 
\begin{equation}\label{eq:spin_eq}
    \frac{d\mathbf{S}}{dt}=\mathbf{\mu}\times\mathbf{B}+\mathbf{d}\times\mathbf{E}+ \mathbf{S}\times \frac{\mathbf{\nabla} \phi}{f_\phi}\,.
\end{equation}

The spin precession of a charged particle has 3 contributions -- the two well-known magnetic dipole and electric dipole moment contributions, and the axion gradient. Any experiment looking for precise measurements of the first two can in principle be used to search for axion mediated forces. 

Crucially for precision precession experiments, the axion gradient is usually an environmental effect which cannot be screened by any kind of magnetic/electric shielding. It is also noteworthy that the axion gradient couples only to spin, and not to orbital angular momentum.  Therefore, any precision experiment may see an effect if the direction of the axion gradient (e.g.~from the earth) is such that the
phase accumulates during the spin coherence time.  

\subsection{Geometry of storage ring experiments and axion forces}

In this section we study how storage rings which are initially designed to measure precisely the spin precession of a charged particle due to an EDM or anomalous magnetic moment (see Eq.~\ref{eq:MDM-EDM}), are also sensitive to axion gradients sourced by matter. As we have seen in the equations above, (\ref{eq:monopole-dipole_Hamiltonian}), (\ref{eq:spin_eq}), an axion gradient behaves as an effective magnetic field. 

In storage ring experiments, a large number of charged, polarized particles perform circular motion under applied electromagnetic fields. It is important to carefully study the direction of the axion gradient relative to the spin and make sure that the effect is not averaged out \cite{Graham:2020kai}.

A magnetic storage ring (e.g. muon $g-2$) has a vertical applied magnetic field, which causes spin precession in the horizontal plane. The effect of a horizontal axion gradient will average to zero over a spin precession cycle, therefore this geometry will be sensitive to a vertical axion gradient, e.g. that sourced by the earth.

A proton storage ring in the all-electric mode has a radial electric field with potentially both clockwise (CW) and counter-clockwise (CCW) rotating beams. On the magic momentum, $p\approx 0.7$ GeV, the spin direction is frozen relative to the momentum of the proton. An electric dipole moment will cause the spin to precess out of the plane. A vertical axion gradient will behave like a fake vertical magnetic field, and cause horizontal precession. Systematics from stray magnetic fields in this case can be mitigated by operating in the hybrid mode \cite{Haciomeroglu:2018nre}. It is also interesting to consider sourcing the axion fields with test masses near the storage ring, which leads to the possibility of choosing the direction of the axion gradient. This allows, for example, a radial axion gradient and also improves the sensitivity to axions with a heavier mass. 

\subsection{Distinguishing axion forces from dipole moments}
It is very interesting to investigate if given a signal in these experiments, we can disentangle its explanation from an axion gradient or other new physics contributions to the anomalous MDM and EDM operators:
\begin{equation}\label{eq:MDM-EDM}
    a_\psi \Bar{\psi}\sigma_{\mu\nu}\psi F^{\mu\nu}\,,\,\,i d_\psi \Bar{\psi}\sigma_{\mu\nu}\gamma_5\psi F^{\mu\nu} \,.
\end{equation}

In a given storage ring experiment there are 3 important \textit{binary} inputs that one can in principle switch individually\footnote{In principle the direction of the spin relative to the momentum can also be chosen in different ways. For simplicity we will assume that the spin and momenta are aligned.}.  These flips can help to cancel systematic errors, but can also be used to distinguish axion gradients from EM effects. First of all, the experiment can be performed with particles and anti-particles. Additionally, we have the rotational direction -- that is, clockwise (CW) and counter-clockwise (CCW). Finally, one can in principle also switch the direction of the electromagnetic fields that keep the charged particle orbiting the ring. Of course, once 2 of the 3 handles are fixed the third one is also uniquely determined for the charged particle to remain in the ring. For example, let us take a magnetic storage ring with muons rotating clockwise. We can switch the orbital direction of the muons and flip the magnetic field at the same time. Alternatively, we can run the experiment with anti-muons also rotating clockwise,  again flipping the sign of $B$. As we will see in the next section, this was indeed the situation in BNL experiment. 

In order to disentangle from EM effects, we need to understand how the axion gradient affects a given spin for the different switches. From Eq.(\ref{eq:spin_eq}) it is clear that, for a fixed axion background, switching the rotational direction will contribute in opposite ways to the spin precession frequency. However, how switching from particles to anti-particles affects the rotation is not immediately clear from that equation and requires a relativistic study. 

To this end we start with the Dirac equation for a charged particle interacting with an axion gradient through derivative dimension-5 operators (see Eq. \ref{eq:lag}):
\begin{equation}\label{eq:Dirac_eq}
    \left(i\gamma^\mu\left(\partial_\mu-ieA_\mu\right)-m+\frac{\partial_\mu \phi}{f_\phi}\gamma^\mu\gamma_5 \right)\psi=0\,.
\end{equation}
It is precisely this equation which, in the non-relativistic limit and taking a static axion background $\partial_t\phi=0$, 
gives the spin equation \ref{eq:spin_eq}.\footnote{The QED+axion Lagrangian leads to vanishing EDM $d_\psi=0$. This may be altered by heavy new physics, at the loop level.} 
We can now use the charge conjugation operator to study the Dirac equation for $\psi^c=C\gamma^0\psi^*$, with $C=i\gamma^2\gamma^0$, which reads:
\begin{equation}\label{eq:Dirac_eq_antiparticle}
\left(i\gamma^\mu\left(\partial_\mu+ieA_\mu\right)-m+\frac{\partial_\mu \phi}{f_\phi}\gamma^\mu\gamma_5 \right)\psi^c=0 \,.
\end{equation}
First of all we see that we recover the standard result -- that is, any EM effect on anti-particles is reversed with respect to particles.
 On the other hand, crucially, the axion gradient does not flip its sign, implying that the effect on the spin of the anti-particle is identical to that on the spin of the particle. 
The fact that for a fixed spin rotational direction particles and antiparticles precess in the same direction in an axion background is guaranteed by the relative sign of the axion dipole term with respect to EM effects in Eqs.(\ref{eq:Dirac_eq}, \ref{eq:Dirac_eq_antiparticle}). 

In some experiments we will be sensitive to the axion gradient from the earth nucleons. This field, obviously, cannot be reversed. Therefore when we perform the measurement of the precession frequency for both spin orientations -- that is, we shift the direction of rotation -- the axion contribution to the spin precession frequency, $\omega_{\text{axion}}$, will present a relative sign for the CW and CCW rotations. This fact is crucial to discriminate from other EM effects such as heavy physics contributions to $(g-2)$, which will equally contribute to the spin precession frequency. In particular we  will see that it has enormous importance for the muon $(g-2)$ experiment.  

\section{Muon storage rings: A new explanation of the $g-2$ muon anomaly}
\label{sec:muon}

The FNAL muon $(g-2)$ experiment \cite{Muong-2:2021ojo}, which is the continuation of BNL experiment \cite{Muong-2:2006rrc}, is a muon storage ring that searches for new physics by measuring the muon magnetic anomaly $a_\mu$ (sometimes called ``anomalous magnetic moment'') with unprecedented precision. The current combined results lead to a $4.2\sigma$ discrepancy with the SM prediction\footnote{See however \cite{Borsanyi:2020mff} for a lattice calculation where the SM prediction is closer to the experimental value. The discrepancy between different approaches to the SM prediction should be clarified soon.}:
\begin{equation}
    a_\mu(\text{EXP})-a_\mu(\text{SM})=(251\pm 59)\times 10^{-11}\,.
\end{equation}

The idea is that muons are stored in the ring until their decay and the spin precession frequency is accurately measured.  The anomalous precession, which determines how the spin rotates faster than the muon momentum, is given by:
\begin{equation}
    \omega_a=\omega_s-\omega_c=a_\mu\frac{e B}{m_\mu}\,.
\end{equation}
The  magnetic anomaly $a_\mu$ usually receives contributions from new physics at the loop level. A SMEFT analysis  reflects that due to the chiral structure of the SM the leading operators that contribute to the anomalous magnetic moment arise at dimension-6, and schematically look like (see \cite{Allwicher:2021jkr} for a recent study):
\begin{equation}\label{eq:anomalous_dim6}
    \Delta a_\mu=a_\mu(\text{EXP})-a_\mu(\text{SM})\sim C_{a_\mu} \frac{m_\mu^2}{\Lambda^2}\,,
\end{equation}
where $C_{a_\mu}$ includes the relevant couplings and a possible loop supression. 

Many models have been proposed in order to explain the BNL and FNAL results \cite{Athron:2021iuf}.
Our proposal is qualitatively different to all the  BSM models studied so far. We assume there is a light axion that couples to muons with the couplings given in Eq.(\ref{eq:lag}) -- that is, a scalar coupling to nucleons and pseudo-scalar coupling to muons. As described above, the axion gradient sourced by the earth will induce spin precession of the stored muons because it has the correct orientation -- it points vertically, in the earth radial direction. Therefore, by virtue of Eq.(\ref{eq:spin_eq}), the axion gradient behaves as an effective magnetic field inducing the same effect as an anomalous  magnetic moment for the muons orbiting around the ring. If the muons are stored for a time $\tau_{coh}$, the precessed angle due to the axion field contribution is given by 
\begin{equation}\label{eq:axion_contribution_g-2}
    \theta_\mu=\tau_{coh}\frac{1}{f_\phi}\nabla \phi\,.
\end{equation}
 
The BNL muon $g-2$ experiment was operated in the  $\mu^+$-mode in the same ring and direction (clockwise from above) as FNAL, also operating with anti-muons and the same magnetic field direction, pointing vertically upwards. In addition, BNL was operated with muons $\mu^-$, keeping the direction and flipping the sign of the magnetic field. Both muons and antimuons gave compatible measurements of $(g-2)_\mu$ \cite{Muong-2:2006rrc}. This is compatible with $a_\mu$ being generated by heavy new physics as in Eq.(\ref{eq:anomalous_dim6}).  Since an axion gradient from the earth cannot be flipped, it is interesting to ask whether the effect from the axion force stays the same or flips sign. From the discussion in section~\ref{sec:setup} in Eq.(\ref{eq:Dirac_eq_antiparticle}), since the spin of particles and antiparticles experience the same torque in an axion background, the axion gradient explanation can accommodate both, BNL ($\mu^-$ and $\mu^+$ modes) and Fermilab $g-2$ results.

\begin{figure*}[t]
	\centering
	\includegraphics[width=0.85\textwidth]{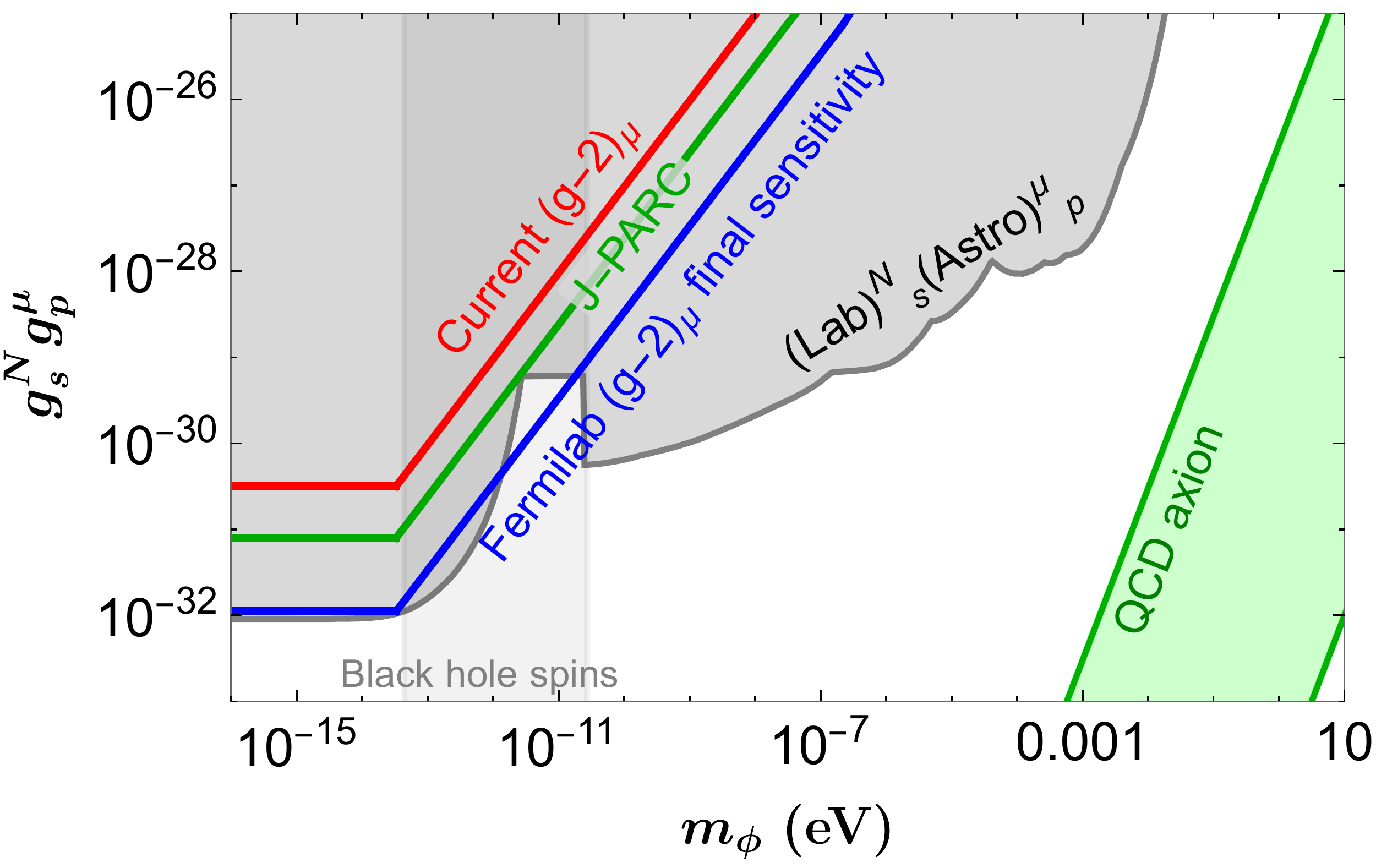}
		\caption{Axion-mediated monopole-dipole forces on muons. The red line corresponds to the values required to explain the $(g-2)_\mu$ anomaly at Fermilab and BNL assuming a signal at the level $\delta \omega_a \approx 4$ rad/s \cite{Muong-2:2006rrc,Muong-2:2021ojo}. Bounds from astrophysics are shown in gray.  The anomaly can be more easily explained in  the region around $m_{\phi}\sim 10^{-12}$ eV, where the new force limits are the weakest. This is still in slight conflict with SN bounds. Note that the stronger bound on a new force with $m_{\phi} \sim 10^{-10}$ eV would also apply at $m_{\phi}\sim 10^{-11.5}$ eV if the force violated the EP maximally. Thus, for this relaxation of limits, we require the new force to obey the EP at the $\sim 1$ percent level, which is reasonable in many models ({\it e.g.} mediation via the higgs). The green line corresponds to the expected sensitivity at J-PARC, which will reach the 0.45 ppm level of precision \cite{Abe:2019thb}. The final sensitivity of Fermilab $(g-2)_\mu$, at the 0.1ppm level which corresponds to $\delta w_a= O(0.1)$ rad/s, is in blue. }

	\label{fig:axion_muon_forces}
\end{figure*}

The axion gradient from the earth nucleons is potentially  an interesting environmental explanation for axion masses below $m_\phi\sim 10^{-10}$ eV, and in particular around $m_\phi\sim 10^{-12}$ eV as shown in Fig.(\ref{fig:axion_muon_forces}). This explanation, however, is in mild tension with astrophysical constraints \cite{Bollig:2020xdr,Caputo:2021rux}. These bounds may be relaxed in some scenarios \cite{DeRocco:2020xdt}.

Very remarkably, it may be possible to distinguish this explanation from other BSM  solutions to the $g-2$ puzzle.
The smoking gun signature that allows to distinguish them is the following. As discussed above, the sign of the axion gradient contribution to the precession frequency should flip by changing the beam rotational direction -- and therefore the spin -- of the muons. This allows a discrimination with respect to other effects such as heavy, new physics contributions to $(g-2)_\mu$, which will deviate equally from the SM prediction in both rotational directions. Therefore we propose to measure  the anomalous precession frequency clockwise, $\omega_a^{cw}$, and counter-clockwise $\omega_a^{ccw}$, appropriately flipping the B field. A non-zero difference:
\begin{equation}
    \Delta \omega_a=\omega_a^{cw}-\omega_a^{ccw}\,,
\end{equation}
is a unique signal of the axion gradient that cannot be accomodated by new heavy physics contributing to $(g-2)_\mu$. Additionally, all the magnetic effects cancelling also implies that the observable $\Delta\omega_a$ is completely free from hadronic uncertainties. 
Although this might require significant modifications to the injector system, we believe there is enough motivation to pursue this direction.

The final senstivity of Fermilab $(g-2)_\mu$ is expected to be at the level of 100 ppb. This corresponds to a sensitivity to the anomalous spin precession frequency, $\omega_a$, around $\delta w_a=O(0.1)$ rad/s and corresponds to the blue line in Fig.\ref{fig:axion_muon_forces}. We see that the Fermilab experiment will be sensitive to new parameter space not constrained by supernova observations.

\subsection{Other muon facilities: J-PARC and muEDM}
There are other planned facilities that aim to measure very precisely the anomalous magnetic dipole moment\footnote{The anomalous magnetic moment might also be determined with ppm sensitivity using muonium spectroscopy in a near future \cite{Delaunay:2021uph}.} and electric dipole moment of the muon. One of them is J-PARC \cite{Abe:2019thb}, which is designed to measure the muon spin precession frequency relative to the momentum frequency, $\omega_a=\omega_s-\omega_c$, with 450 ppb accuracy. Given the values for the B field and other parameters, this corresponds to a sensitivity to the spin precession frequency of order $\delta w_a \approx 1$ rad/s (see Fig.\ref{fig:axion_muon_forces}). 

Another interesting facility that will carry out muon spin precession experiments is muEDM \cite{Adelmann:2021udj,Sakurai:2022tbk}. The proposed experiment aims to look for the muon EDM with unprecedented precision in an electric storage ring, with the spin ``frozen" so that there is no precession of the spin in the plane of the ring on the magic momentum.
The employed electric field is in the radial direction, so any spin precession is observed out of the plane of rotation. The effect of an EDM for the muon will be measured by counting positrons resulting from its decay -- an EDM would generate an up/down asymmetry. Due to this configuration, however, muEDM will not be sensitive to a vertical axion gradient from the ground -- the one that may explain the muon $(g-2)$ anomaly -- because it generates a spin precession on the horizontal plane.

\section{Proton storage rings}
\label{sec:protons}
The proton EDM storage ring is a proposal that is currently being designed and aims to measure the EDM of the proton \cite{Alexander:2022rmq,CPEDM:2019nwp}. The planned sensitivity is at the level of $d_p\lesssim 10^{-29}$ e$\cdot$cm, allowing to improve current bounds by several orders of magnitude. 

In the all-electric mode, where both focusing and bending of the beams are generated by electric fields, there is a radial electric field $\mathbf{E_r}$, and the beam is tuned to the magic momentum with the spin frozen in the horizontal plane. A non-zero proton EDM will cause the spin to precess around the applied radial electric field, therefore in the vertical direction. A hybrid mode \cite{Haciomeroglu:2018nre,PhysRevD.105.032001}, which still uses electric bending but employs magnetic focusing of the beams has been also proposed to address systematic uncertainties with stray radial magnetic fields. In either configuration the background has to be kept under control to extraordinary levels of precision. This requires different techniques such as using state-of-the art magnetic shielding and precise beam position monitor stations that measure the split of counter-rotating beams due to background magnetic fields. This may in principle be enough to reach the expected sensitivity. 

Even though the current experimental design is not optimized for an axion gradient from the earth, the high precision goal for proton EDM makes this a promising target even with some decreased sensitivity. Additionally, as we will see, it may be possible to detect axion gradients sourced from test masses in the lab. 
\subsection{Axion gradient from the earth}
As mentioned above, the axion gradient from earth produces an effective $B$ field in the vertical direction, causing the spin to precess in the horizontal plane. Since the experiment is in principle designed to look for a vertical polarization, 
this might be a difficult measurement to make.

One source of systematics would be that the axion gradient would be degenerate with a stray vertical magnetic field. However, this systematic effect can be mitigated. The proton ring can accommodate counter rotating beams at the same time, which would split in the presence of a vertical magnetic field component. The axion gradient does not affect the proton trajectories since it only couples to spin. Thus, the two counter rotating beams give us an extra handle to distinguish axion gradients from magnetic field backgrounds.

The sensitivity to horizontal spin precession may be somewhat lower than that for the vertical precession. As an optimistic guess, we estimate the same precision on the horizontal precession.
The estimated reach, once the experiment achieves its design sensitivity $d\lesssim 10^{-29}$ e.cm corresponds to a spin precession rate of $\delta \omega=10^{-9}$ rad/s and is given in Fig. \ref{fig:axion_proton_forces}. We note that the projected sensitivity is 4-5 orders of magnitude stronger than current constraints on very light axions. So even if the ultimate sensitivity is much worse, proton rings may still offer the strongest constraint on axion-nucleon forces in this parameter range. It will be important to better understand with a dedicated study the relevant systematics and the achievable sensitivity to the signal.

\subsection{Axion gradients from sources}
The vertical precession is not sensitive to the earth gradient, but can be sensitive to axion gradients from localized sources. This can be the dominant effect also 
for shorter range forces. 

\begin{figure*}[t]
	\centering
	\includegraphics[width=0.85\textwidth]{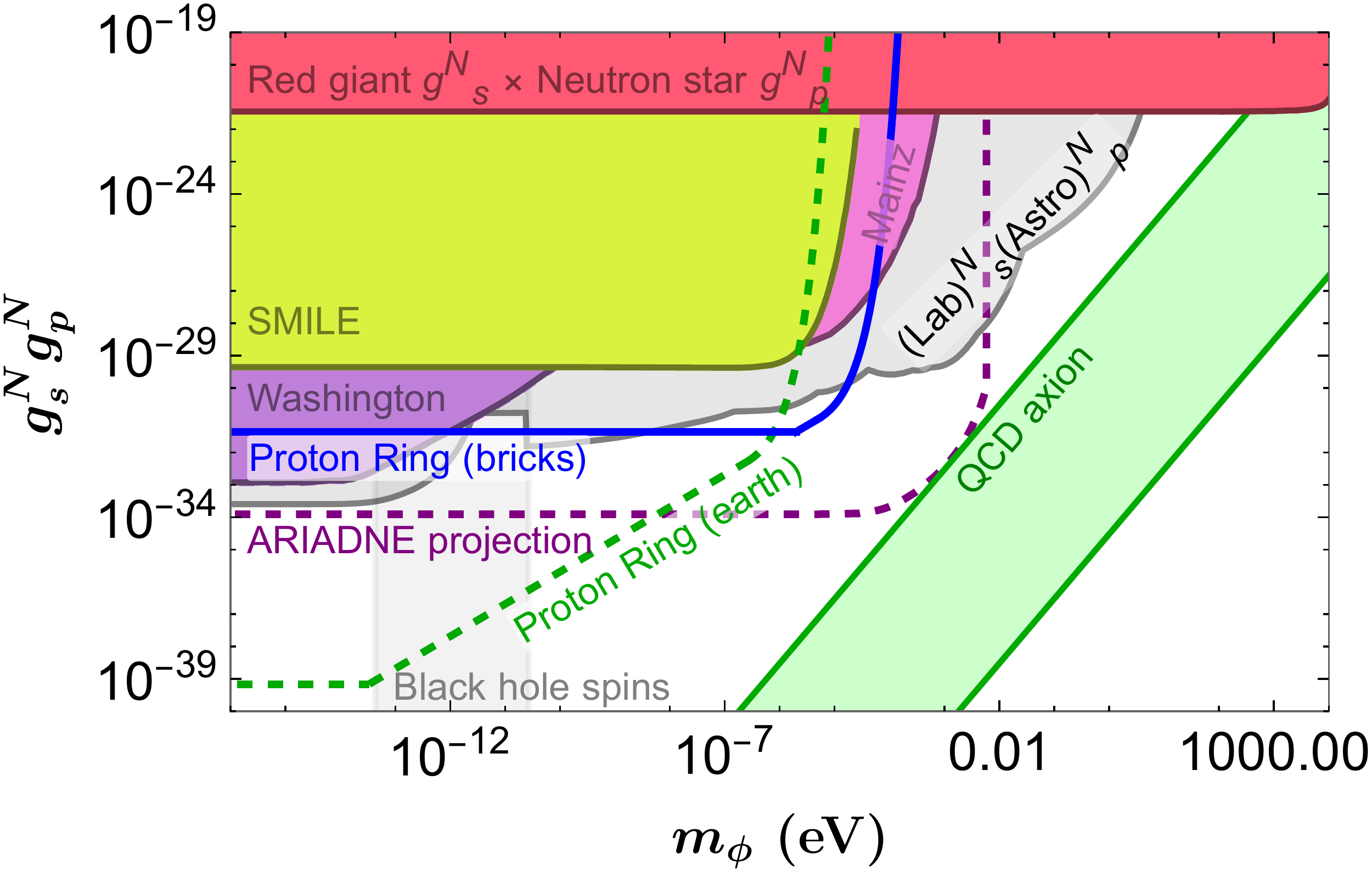}
		\caption{Expected sensitivity to axion forces in the proton storage ring experiment. The blue line corresponds to cubic lead bricks of 10cm size around the ring at a distance around 10cm from the beams generating spin precession out of the plane (EDM-like signature). We consider the conservative case of covering only $O$(10\%) of the ring with bricks. This fraction is an important parameter, as the reach increases linearly with it. The green dashed line corresponds to the limits due to the axion field from the ground nucleons (spin precession on the ring plane). It is assumed that the ring is located around 150 cm above the ground.  The combination of both configurations gives the strongest bounds to monopole-dipole forces on nucleons, beating astrophysics and existing laboratory bounds for any mass below $m_{\phi}=10^{-5}$ eV. Bounds adapted from \cite{OHare:2020wah}.}
	\label{fig:axion_proton_forces}
\end{figure*}

In Fig.\ref{fig:axion_proton_forces} we have estimated the reach assuming the presence of cubic bricks acting as axion field sources around the ring. If placed inside or outside the beampipe, on the ring plane, the relevant axion gradient will point in the radial direction, producing the spin precession in the vertical direction. Note that with this geometry we can benefit from the counter-rotating (CR) beams -- such axion-induced precession has the `EDM-like' signature, where the direction of the vertical precession for the CR beams are opposite.

The proton ring is expected to have a total length around 800 meters (see \cite{Alexander:2022rmq} for details), so covering the whole circumference may be challenging. Due to technical reasons there may be also some parts of the ring where no brick can be placed, for example close to the injectors. We have assumed that we have a $O(0.1)$ fraction of the ring covered with bricks made of lead to maximize the number density of nucleons, $n_{\text{lead}}=6.8\times 10^{24}\text{ cm}^{-3}$ (tungsten bricks may give a extra factor of $\sim2$ for the reach, due to the higher density). This indeed seems reasonable, as it only requires using about 1m$^3$ of lead (approx 10 tons) uniformly distributed in (10cm)$^3$ bricks around the ring. Interestingly, a proton EDM ring using sourced axions from test masses will be able to beat astrophysical bounds in the mass range $10^{-8.5}\text{ eV }\lesssim m_\phi\lesssim 10^{-4.5}$ eV. This bound is also expected to be around 2 orders of magnitude better than existing lab bounds in a wide range of masses. 

\section{Quantum electron cyclotrons and the $(g-2)_e$}
\label{sec:electrons}
Precise spin precession experiments with electrons are the another possibility that storage rings may offer. 
In particular, experiments with electrons in cylindrical Penning traps \cite{Hanneke:2008tm,Hanneke:2010au} seem to be well-suited to search for axion mediated forces from earth nucleons. The trap consists of a 1 electron quantum cyclotron where a magnetic field $\mathbf{B}=B\hat{z}$ is employed together with an electrostatic quadrupole potential $V=z^2-\rho^2/2$, which confines the electron axially. As for the muon case, this requires observing a DC signal from an axion gradient in the vertical direction which contributes to the anomalous precession frequency on the quantum cyclotron plane.  Being the most accurate measurement of $(g-2)_e$, where both the cyclotron and anomalous frequency are very precisely measured at the level of 1 part in $10^{10}$, it is interesting to estimate the sensitivity to axion-mediated monopole-dipole forces. 

\begin{figure*}[t]
	\centering
	\includegraphics[width=0.85\textwidth]
 {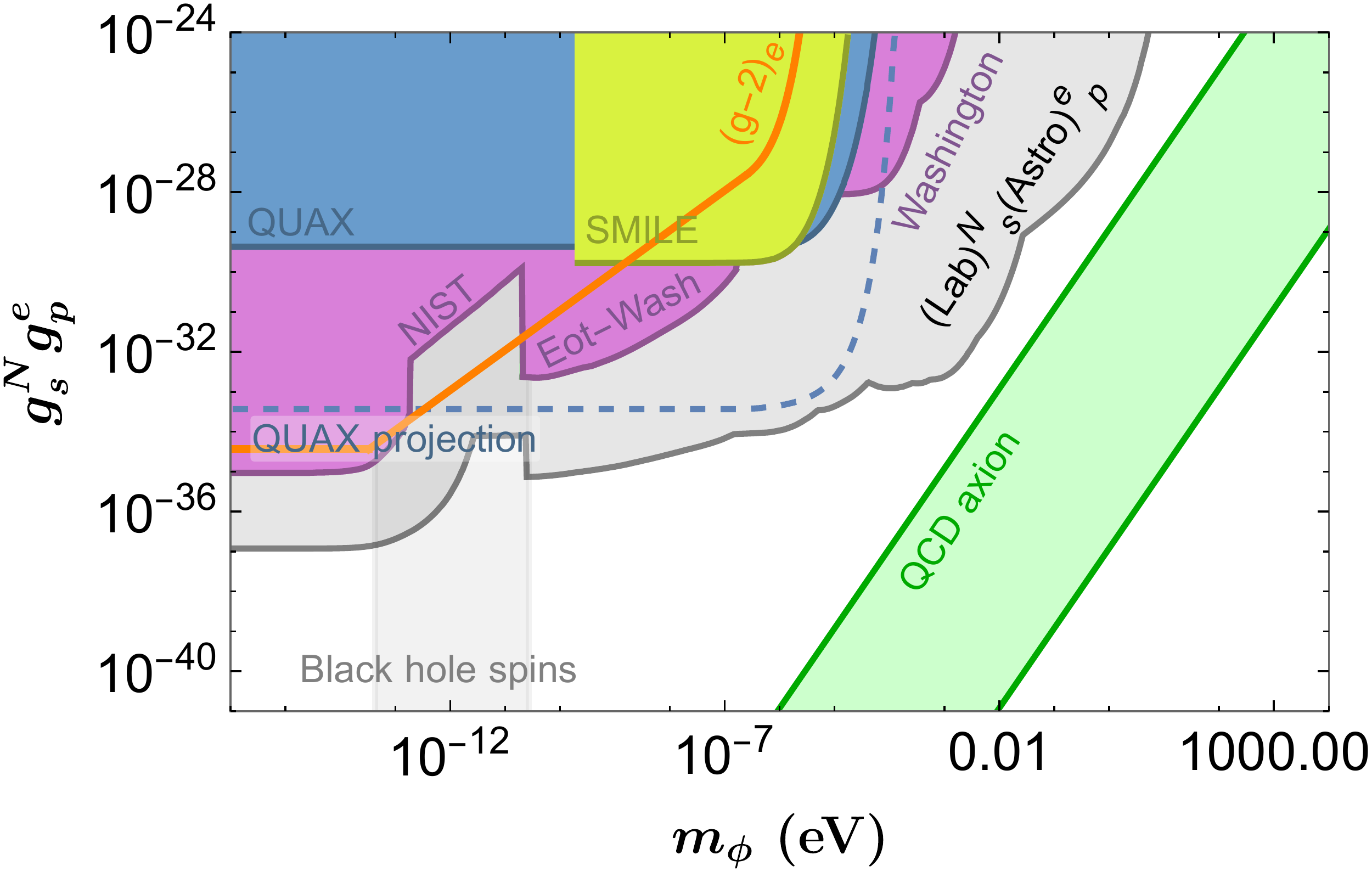}

		\caption{Sensitivity estimate to axion forces on electrons from earth nucleons in the $(g-2)_e$ experiment \cite{Hanneke:2008tm,Hanneke:2010au} (orange line).  It is assumed that the Penning trap is located around $O(1)$ meter above the ground. For completeness, other laboratory bounds and expected sentivities are shown. Bounds adapted from \cite{OHare:2020wah}.}

	\label{fig:axion_electron_forces}
\end{figure*}

In the Penning trap the electron orbits very fast, with a cyclotron frequency of $\nu_c\sim 149$ GHz, due to a strong magnetic field. Because of the non-zero value of the anomalous magnetic moment, the spin precesses faster than the momentum. Their difference is precisely measured to be $\nu_s-\nu_c=\nu_a\sim 173$ MHz. The anomalous precession frequency is measured with a precision better than one part in $10^{10}$ \cite{Hanneke:2008tm,Hanneke:2010au}, implying that their sensitivity to the electron frequency is around:
\begin{equation}
    \delta\omega_a \sim  O(0.1) \text{ rad/s}\,.
\end{equation}
While the quantum cyclotron sets by far the best measurement of the electron anomalous magnetic moment, its expected sensitivity is not enough to beat existing bounds to the monopole-dipole axion forces in the mass range $m_\phi>2\times 10^{-11}$ eV. However, the experiment could be used to set the best bounds for masses in the range $2\times 10^{-13} \text{ eV}<m_\phi<2\times 10^{-11}$ eV. See Fig. \ref{fig:axion_electron_forces}.

\section{Conclusion}
\label{sec:conclusion}
Storage rings provide some of the most precise experimental tests in fundamental physics by building up precession of spin over a long time. They provide, since many years ago, the best constraints on electric and magnetic dipole moments of particles.

In this paper we have shown that these experiments are also sensitive to gradients induced by light spinless fields, axions. Specifically, the earth can source an axion gradient which can produce a muon spin precession almost as big as the current discrepancy between theoretical calculations and experiment. Future, higher precision experiments will unambiguously probe new parameter space, and experimental progress in this direction is not contingent on improvement of difficult theoretical calculations. The signal also motivates a modification to the injector in the Fermilab experiment with a beam rotating in the opposite direction, which can help distinguish the axion gradient effect from other new physics contributions to the muon $g-2$. The observable that disentangles them is, remarkably, free from hadronic uncertainties.

Depending on the geometry, proton and electron storage rings can also probe new regions of axion-fermion couplings. More experimental work needs to be done to ascertain systematics in this new configuration. The axion gradient effect can be distinguished from other signals depending on the geometry of the experiments. As the next generation of storage rings are being planned, this provides a vital physics case input to the design of these experiments.
\\\\
\textit{Note added:} While we were finishing the writing of this manuscript reference \cite{Davoudiasl:2022gdg} appeared, partially overlapping with our work. However there are fundamental differences.  Regarding the explanation of the muon anomaly, for example, we show that the tension with astrophysics is minimized for axion masses around $10^{-12}\text{ eV }\lesssim m_\phi\lesssim 10^{-11}$ eV. We also noted that it is possible to further confirm the effect and discriminate from other heavy new physics signals just by performing the $(g-2)_\mu$ measurement in the opposite rotational direction, with no need for constructing a new experiment in a different location/astronomical body. Finally, in the proton ring case, we note that using test masses of a dense material may probe shorter range forces mediated by heavier axions. 
\\\\
\textit{Note added II}: While this manuscript was being reviewed two relevant results appeared in the literature in relation to the muon $(g-2)_\mu$. The first one is the CMD-3 measurement of the $e^+e^-\rightarrow \pi^+\pi^-$ cross-section \cite{CMD-3:2023alj}, which seems to reduce the tension between theory and the experiment. The second is a new lattice calculation which also reduces
the tension  \cite{Blum:2023vlm}.  In the case of the CMD-3 result, there seems to be a disagreement with
other existing measurements, including with the previous measurement by CMD-2.  Regarding \cite{Blum:2023vlm}, it will be very interesting to see the evolution of the lattice results and their impact on the need for new physics for muon $(g-2)_\mu$. 
We would like to emphasize that our work motivates pushing storage ring precession experiments as probes of axion-mediated spin-dependent forces independently of the muon $(g-2)_\mu$ anomaly.

\section*{Acknowledgments}
D.~E.~K and S.R.~are supported in part by the U.S.~National Science Foundation (NSF) under Grant No.~PHY-1818899.  
This work was supported by the U.S.~Department of Energy (DOE), Office of Science, National Quantum Information Science Research Centers, Superconducting Quantum Materials and Systems Center (SQMS) under contract No.~DE-AC02-07CH11359. S.R.~is also supported by the DOE under a QuantISED grant for MAGIS, and the Simons Investigator Award No.~827042.  P.A. is supported by the STFC under Grant No. ST/T000864/1. O.~K. is supported by the U.S. DOE under contract no. DE-SC0021616.
M.R. would like to thank the Physics department at JHU where part of this project was carried out. 


\providecommand{\href}[2]{#2}\begingroup\raggedright\endgroup

\end{document}